# Quantum antenna as an open system: strong antenna coupling with photonic reservoir

**Alexei Komarov and Gregory Slepyan \***

School of Electrical Engineering, Tel Aviv University,
Tel Aviv 39040, Israel; al3x3i.k@gmail.com
\* Correspondence: slepyan@post.tau.ac.il; Tel.: +972-54-737-89-17

**Abstract:** We proposed the general concept of quantum antenna in the strong coupling regime. It is based on the theory of open quantum systems. The antenna emission to the space is considered as an interaction with the thermal photonic reservoir. For modeling of the antenna dynamics is formulated a master equation with the correspondent Lindblad super-operators as the radiation terms. It is shown that strong coupling dramatically changes the radiation pattern of antenna. The total power pattern splits to three partial components; each corresponds to the spectral line of Mollow triplet. We analyzed the dependence of splitting from the length of antenna, shift of the phase, and Rabi-frequency. The predicted effect opens a way for implementation of multi-beam electrically tunable antennas, potentially useful in different nano-devices.

**Keywords:** Quantum antennas, Rabi-oscillations; Photonic reservoir, Master equations,

## 1.Introduction

Traditionally, the field of quantum optics deals with the control of quantum statistical correlations of light using elements such as interferometers, cavities, photonic crystals, beam splitters, etc [1]. Purcell effect [2] and Dicke's superradiance effect [3] are two of the most intriguing effects in quantum optics. Purcell predicted in 1946 the dependence of spontaneous emission rate on the environment of the light source. The effect is based on the ability of the specific environment to control the photonic density of states at the source position. Dicke introduced in 1954 the concept of enhanced spontaneous emission by an optically small ensemble of identical two-level atoms collectively interacting via a common electromagnetic field [3]. This effect involves the collective Dike's states which turn out to be highly entangled. The problem becomes even more fascinating if instead of a small atomic cloud, the size of the system becomes comparable with the wavelength. In 2006, Scully with co-workers focused on the problem of a single photon stored in a large cloud of atoms, which exists in a collective *N*-atom state [4-9]. Photon is shared among the atoms, thus one of them is excited, but we do not know which one. When the atoms in various positions are excited at different moments in time (Dike's states timing), the usual picture of collective emission, which treats each atom as a tiny antenna, is no longer valid. In this case, peculiar features of the directional and temporal characteristics of the cooperatively emitted radiation have been predicted. In particular, a single photon absorption by the cloud of *N* atoms is followed by the spontaneous emission in the same direction. These investigations gave birth to what is known today as "correlated spontaneous emission" and opened one of the possible ways for quantum antennas design.

Rapid progress in communication and radar technologies for the radio frequency ranges (from microwaves to terahertz) led to the development of various types of antennas that control electromagnetic fields on the wavelength scale [10-13]. The progress in modern nanotechnologies brings about the general trend of the transferring principles of radio-communication to the optical range. Thus, optical antennas became a promising tool for manipulating and controlling optical radiation [10-13]. In general, the classical antenna is defined as a device transforming the energy of coupled near field to the free far field (transmitting antennas) and vice versa (receiving antennas) [14]. For many practical applications, the emitted field should be directive: the main part of its energy is concentrated in the narrow main lobe. For highly-efficient antennas the "near-far zone" coupling (defined, for example, by radiation resistance [14]) should be sufficiently strong.

Classical antenna concepts [14] with respect to the quantum antennas need a fundamental reconsideration. The reason for this is that an emitting elements and EM-field generators (for transmitting antennas) or detectors (for receiving antennas) are completely constructively separated. In contrast with classical case, the antenna emitters cannot be considered independently from antenna elements. The radiation antenna properties are not separable from the source's and detector's origin. It becomes impossible to speak about antennas as passive linear elements. The antennas become nonlinear and active (for recent reviews – see [15,16]). Antenna nonlinearities in their typical implementations are so strong, that consideration of nonlinearity as a weak perturbation breaks down. Strong and even ultra-strong coupling of emitter and EM-field became a reality [17]. As a result, such fundamental effects as spontaneous emission [18,19], Rabi-oscillations [20,21] and solitons [22,23] become the basis principles for quantum nanoantennas. The different types of nanostructures (plasmonic nanowires [24], quantum dots [25], carbon nanotubes [26-28] etc.) have been considered as promising materials for their implementation. As a result, antennas open a new ways for control by the directed quantum emission and the correlation light properties [29]. In a lot of cases the effect of non-reciprocity manifests itself in quantum antennas [30,31], which leads to the difference of the same antenna properties in transmitting and receiving regimes without using the ordinary types of reciprocal materials. On the other hand, the quantum origin leads to some fundamental limitations, which are not installation-specific classical antennas (for example, Heisenberg uncertainty principle [32]).

In this paper we focus on the theory of quantum-optical antenna with fermionic type of excitation. Our description is based on the such general principles of quantum physics as a theory of open systems [1,33] with their bridging to the classical antenna theory [14]. The general principle of open system theory is based on the separation to the actual system and evanescent space (reservoir). The last one assumed to be so large that its interaction with the actual system doesn't change its quantum state. We consider the quantum antenna as a radiating actual system and a surrounding space in which radiation takes place – as a reservoir. The interaction between the actual system and reservoir leads to the losses in it. Therefore, the theory of open systems is an effective general technique for description of losses of different physical origin may be applied to the problem of antenna emission. Antenna's emission have been considered as a loss of energy of the source placed in space (this is the reason to speak about radiative losses). The influence of antennas on the state of surrounding area is negligibly small. This is the reason to consider the last one as photonic reservoir. The model of photonic reservoir in the thermal state corresponds to the transmitting antenna (the antenna evanescent appears in thermally equilibrium state and produces the thermal noise). Our considerations have yielded the fundamental properties of quantum antennas, which have no analogs in the classical ones and provide applications in future.

The paper is organized as follows: in the Section 2 we formulate the model of quantum antenna as a quantum system, formulate the basis Hamiltonian and correspondent master equation. In Section 3 we study dynamics of quantum antenna basing on the approximate analytical solutions in the regimes of weak and strong coupling. The power radiative patterns of quantum antenna are studied in Section 4. Conclusions, main results and their promising applications are formulated in Section 5.

## 2. The model of quantum antenna as an open system

The general model of the antenna is shown on Figure 1. We consider the antenna as a quantum dot placed inside the quantum wire, which manifests itself as a waveguide. (Figure 1a). Such type of devices was implemented experimentally at the form of tapered InP nanowire waveguide containing a single InAsP quantum dot [34-36]. Quantum dot is perfectly positioned on-axis of InP nanowire waveguides, where the emission is efficiently coupled to the fundamental waveguide mode.

From theoretical point of view this system may be considered as a defect of a crystal lattice. Therefore, the general model of the bulk crystal defects, which is based on Wannier functions [37], may be used for the antenna analysis. Similar [37], we define Wannier functions $a_n(x - x_j)$ mutually coupled with Bloch functions $\psi_{nh}(x)$ via relations

$$a_n(x - x_j) = \frac{1}{\sqrt{N}} \sum_h e^{-ihx_j} \psi_{nh}(x) \qquad (1)$$

$$\psi_{nh}(x) = \frac{1}{\sqrt{N}} \sum_j e^{ihx_j} a_n(x - x_j) \qquad (2)$$

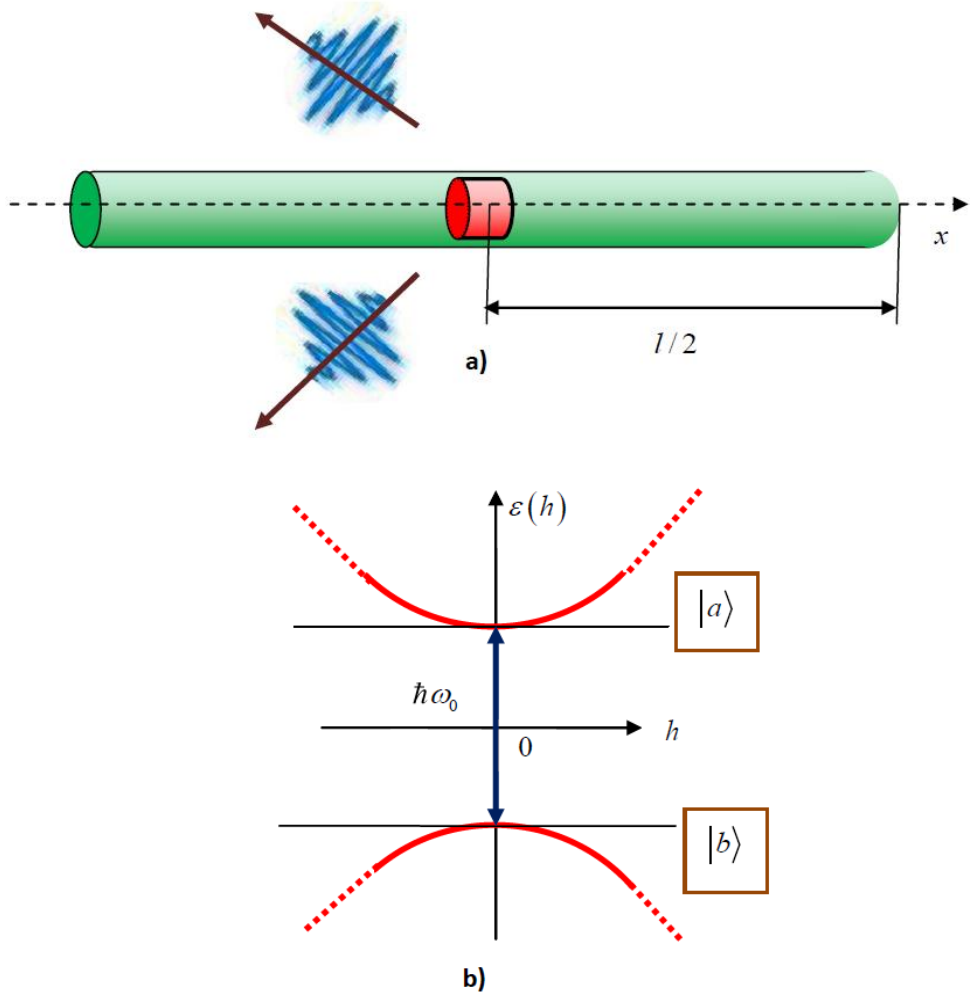

**Figure 1**. General illustration of the system used as a model of quantum antenna in the strong coupling regime. **(a)** The antenna configuration. The semiconductor InAsP quantum dot (red colored) positioned inside the semiconductor InP quantum wire (green colored). The quantum dot is coupled with quantum wire via quantum tunneling through the potential barriers. **(b)** The energy spectrum of the shown system in the case of EM-field absence. It consists out of two zones, valence and conductive, which are separated by the energy gap. The dominative support to the energy of quantum transitions is given by the small region near the band minimum at the zone center and is approximately equal to $\hbar\omega_0$.

where $n$ is a number of zone, $h$ is quasimomentum of the charge. The Wannier function $a_n(x - x_j)$ is strongly localized in the vicinity of the atom at the point $x = x_j$. The arbitrary wave-function is given by $\psi(x) = N^{-1/2} \sum_{n,j} f_n(x_j) a_n(x - x_j)$, where the envelope $f_n(x)$ in the continuous limit satisfies Schrodinger equation

$$\left[ -\left( \frac{\hbar^2}{2m_{eff}} \right)^2 \frac{\partial^2}{\partial x^2} + U(x) \right] f(x) = \left[ \varepsilon - \varepsilon_c(0) \right] f(x) \qquad (3)$$

where $U(x)$ is the potential, averaged with respect to the crystal lattice, $m_{eff}$ and $\varepsilon$ are the effective mass and energy of charge carrier, respectively.

The Bloch function may be presented as $\psi_h(x) = e^{ihx} u_h(x)$, where $u_h(x)$ is a periodic function; $u_h(x+a) = u_h(x)$. The Wannier and Bloch states are satisfy uncertainty principle [37] in the form $\Delta h \cdot \Delta x \cong 1$. It means, that for rather long antenna (large $\Delta x$) the main support to the wave-function is defined by the small region near the band minimum at the zone center. As a result, we obtain the approximate presentation

$$\psi(x) = \frac{1}{\sqrt{N}} \sum_j f(x_j) \sum_h e^{-ih(x-x_i)} u_h(x) \approx$$
$$\frac{1}{\sqrt{N}} u_0(x) \sum_j f(x_j) \sum_h e^{-ih(x-x_i)} = \qquad (4)$$
$$\frac{1}{\sqrt{N}} u_0(x) \sum_j f(x_j) \delta(x-x_i) = u_0(x) f(x)$$

Equation (4) shows, that the quantum properties of antenna may be modeled by the two-level artificial atom with Fermionic quantum states $|a\rangle, |b\rangle$ separated by the energy $\hbar \omega_0$ with the same spatial envelope $f(x)$ (Fig 1b).

We assume the length of antenna to be comparable with the wavelength. The area of quantum confinement is comparable with the wavelength. It makes the retardation of EM-field in its interaction with quantum emitter to be essential. The antenna is driven by the classical external field in the regime of arbitrary coupling. In particular, an essential role in the formation of radiation is played by the Rabi-oscillations produced by the antenna feeding. Therefore, its interaction with EM-field cannot be considered by different types of perturbation theory.

The total Hamiltonian of antenna in the EM-field within the bounds of given model is $\hat{H} = \hat{H}_0 + \hat{V}$, where

$$\hat{H}_0 = \frac{\hbar \omega_0}{2} \left( |a\rangle\langle a| - |b\rangle\langle b| \right) \qquad (5)$$

$\hat{H}_0$ is Hamiltonian of free antenna and

$$\hat{V} = -\frac{\hbar \Omega_R}{2} \left( |b\rangle\langle a| e^{i\omega t} + |a\rangle\langle b| e^{-i\omega t} \right) \qquad (6)$$

is the interaction Hamiltonian with the external field. We will limit our consideration by the strong coupling regime and not touch ultra-strong one [17], thus Hamiltonian (2) is written in rotating-wave approximation [1]. The states $|a\rangle, |b\rangle$ are excited and ground states of antenna without field, $\omega_0, \omega$ are the frequencies of optical transition and driven field, respectively, $\Omega_R$ is the Rabi-frequency.

The radiation of quantum antenna into photonic reservoir is described in rotating-wave approximation [1] by Hamiltonian

$$\hat{H}_{int} = \hbar \sum_{\mathbf{k}} \frac{g_{\mathbf{k}}}{l} \int_{-l/2}^{l/2} e^{-i(\mathbf{k}\cdot\mathbf{e})x} \left[ \hat{b}_{\mathbf{k}} \hat{\sigma}^+(x) e^{i(\omega - \omega_{\mathbf{k}})t} + \text{H.c.} \right] dx \qquad (7)$$

where $g_{\mathbf{k}} = d_{ab} \cos(\mathbf{e}\cdot\mathbf{k}) \sqrt{\omega_{\mathbf{k}}/2\hbar\varepsilon_0 V}$ is a coupling factor of k-th EM-field mode of reservoir with antenna, $d_{ab}$ is the dipole moment, $\hat{\sigma}^+ = |b\rangle\langle a|$ is the Fermionic-type lowering operator of antenna excitation, $\hat{b}_\mathbf{k}$ is a creation operator of photon in the k-th mode, $\mathbf{e}$ is the unit vector along the antenna axis, $\omega_\mathbf{k}, \mathbf{k}$ are the frequency and wave-vector of k-th mode, $V$ is the normalization volume.

We will use the eigenfunctions of Hamiltonian $\hat{H} = \hat{H}_0 + \hat{V}$ as a basis functions. For this we transform all operators following the substitution $\hat{A} \Rightarrow \hat{U}\hat{A}\hat{U}^+$, where $\hat{U}(t) = \begin{pmatrix} e^{i\omega t/2} & 0 \\ 0 & e^{-i\omega t/2} \end{pmatrix}$ is the transformation operator. The Hamiltonian components in this case are transformed following $\hat{H}_0 \Rightarrow \hbar\Delta(|a\rangle\langle a| - |b\rangle\langle b|)/2$, $\hat{V} \Rightarrow -\hbar\Omega_R(|b\rangle\langle a| + |a\rangle\langle b|)/2$, where $\Delta = \omega_0 - \omega$ is the frequency detuning. The eigenmodes are defined by relation $\hat{H}|\Psi_\alpha\rangle = \nu_\alpha |\Psi_\alpha\rangle$, $\alpha = 1,2$. The eigenvalues are given by $\nu_\alpha = \pm\nu = \pm\sqrt{\Delta^2 + \Omega_R^2}/2$, with corresponding eigenmodes

$$|\Psi_1\rangle = Ce^{i\nu t}(g|a\rangle + |b\rangle) \tag{8}$$

$$|\Psi_2\rangle = Ce^{-i\nu t}(|a\rangle - g|b\rangle) \tag{9}$$

where $C = 1/\sqrt{1+g^2}$ is normalization factor, $g = \Omega_R/(\Delta + 2\nu)$. The eigenmodes (8), (9) are orthonormal: $\langle\Psi_\alpha|\Psi_\beta\rangle = \delta_{\alpha\beta}$. This states describe the interaction of high and down states via Rabi-oscillations with frequency $\nu$.

The state of emitting antenna is characterized by the $2\times 2$ matrix of density $\rho$, which satisfies the master equation

$$\dot{\rho} = -\frac{i}{\hbar}[\hat{H}, \rho] + \hat{\Im}(t) \tag{10}$$

The last term is the Lindblad super-operator, which models antenna emission as its coupling with photonic thermal reservoir. It reads in its conventional form

$$\hat{\Im}(t) = -\frac{1}{l}\int_{-l/2}^{l/2}\int_{-l/2}^{l/2} dx dx' f(x) f^*(x') \times \\ \left\{\sum_\mathbf{k} \frac{g_\mathbf{k}^2}{\hbar} e^{i\mathbf{k}\cdot\mathbf{e}(x-x')} \int_0^t [\hat{\sigma}^+\hat{\sigma}^-\rho(t') - \hat{\sigma}^-\rho(t')\hat{\sigma}^+] e^{i(\omega-\omega_\mathbf{k})(t-t')} dt' + H.c.\right\} \tag{11}$$

The next step of simplification consists in the standard transformation from summation over $\mathbf{k}$ to the frequency integration [1] using the replacement

$$\sum_\mathbf{k}(...) \Rightarrow 2\frac{V}{(2\pi)^3}\int_0^{2\pi} d\varphi \int_0^\infty dk \cdot k^2 \int_0^\pi d\theta \sin\theta (...) \tag{12}$$

where azimuthal integration have been carried out and the new variable of integration $k = \omega_\mathbf{k}/c$ have been used. As a result, we obtain

$$\hat{\Im}(t) \approx \frac{d_{ab}^2}{(2\pi)^2 \hbar \varepsilon_0 c^3} \times$$

$$\left\{ \int_0^\pi \sin\theta \cos^2\theta \int_0^\infty \omega_{\mathbf{k}}^3 |F(\omega_{\mathbf{k}};\theta)|^2 \int_0^t e^{i(\omega-\omega_{\mathbf{k}})(t-t')} \left[ \hat{\sigma}^+ \hat{\sigma}^- \rho(t') - \hat{\sigma}^- \rho(t') \hat{\sigma}^+ \right] d\omega_{\mathbf{k}} d\theta dt' + H.c. \right\} \quad (13)$$

where

$$F(\theta,\omega_{\mathbf{k}}) = \frac{1}{l} \int_{-l/2}^{l/2} f(x) e^{-i\frac{\omega_{\mathbf{k}}}{c} x \cos\theta} dx \quad (14)$$

The main support to the time integral is given by the narrow vicinity of frequency $\omega_{\mathbf{k}} \approx \omega$. It allow to use the approximation $\omega_{\mathbf{k}}^3 |F(\theta,\omega_{\mathbf{k}})|^2 \approx \omega^3 |F(\theta,\omega)|^2$ and integrate with $\int_0^t e^{i(\omega-\omega_{\mathbf{k}})(t-t')} d\omega_{\mathbf{k}} \approx \int_{-\infty}^\infty e^{i(\omega-\omega_{\mathbf{k}})(t-t')} d\omega_{\mathbf{k}} = 2\pi \delta(t-t')$. Thus, we have

$$\hat{\Im}(t) \approx -\frac{\Gamma(\omega)}{2} \left[ \hat{\sigma}^+ \hat{\sigma}^- \rho(t) - 2\hat{\sigma}^- \rho(t) \hat{\sigma}^+ + \rho(t) \hat{\sigma}^- \hat{\sigma}^+ \right] \quad (15)$$

with

$$\Gamma(\omega) = \frac{d_{ab}^2 \omega^3}{\pi \varepsilon_0 \hbar c^3} \int_0^\pi \sin\theta \cos^2\theta |F(\theta,\omega)|^2 d\theta \quad (16)$$

The relaxation parameter $\Gamma(\omega)$ is a spontaneous emission frequency. Its difference with Weisskopf – Wigner result for individual atom [1] consists in the special interference dictated by the relative phase shift of different EM-modes over antenna axis. As a result, its frequency dependence doesn't add up to $O(\omega^3)$ as for individual atom.

The presentation of operator $\hat{\sigma}^+ = |b\rangle\langle a|$ at the basis (8),(9) reads

$$\hat{\sigma}^+(x,t) = \sum_{\alpha,\beta=1,2} \kappa_{\alpha\beta} f^2(x) |\Psi_\alpha\rangle\langle\Psi_\beta| e^{2i\nu t} \quad (17)$$

where coefficients $\kappa_{\alpha\beta}$ are elements of matrix $\kappa = C^2 \begin{pmatrix} g & 1 \\ -g^2 & -g \end{pmatrix}$. The equations of matrix of density elements at the basis (8), (9) are

$$\dot{\rho}_{11} = -\frac{1}{2} \left\{ 2\Gamma(\omega) C^4 (1+g^4) \rho_{11} - 2\Gamma(\omega) C^4 g^4 + gC^4 \left[ \Gamma(\omega+2\nu) - \Gamma(\omega-2\nu) \right] (\rho_{12} + \rho_{21}) \right\} \quad (18)$$

$$\dot{\rho}_{12} = -i\left(2\nu - i\frac{1}{2}\gamma_{12}\right)\rho_{12} - \frac{\gamma_{21}}{2}\rho_{21} + \frac{\gamma_{22}}{2} + \frac{(\gamma_{11}-\gamma_{22})}{2}\rho_{11} \quad (19)$$

$$\dot{\rho}_{21} = i\left(2\nu + i\frac{1}{2}\gamma_{12}\right)\rho_{21} - \frac{\gamma_{21}}{2}\rho_{12} + \frac{\gamma_{22}}{2} + \frac{(\gamma_{11}-\gamma_{22})}{2}\rho_{11} \quad (20)$$

The relaxation parameters are obtained from rather long, but trivial calculations. They are given by

$$\gamma_{12} = C^2\left[\left(2-C^2\right)\Gamma(\omega+2\nu) + g^2\left(1+C^2\right)\Gamma(\omega-2\nu)\right] \tag{21a}$$

$$\gamma_{21} = g^2 C^4\left[\Gamma(\omega+2\nu) - \Gamma(\omega-2\nu)\right] \tag{21b}$$

$$\gamma_{22} = \left[gC^2\left(2-C^2\right) + g^3 C^4\right]\Gamma(\omega) \tag{21c}$$

$$\gamma_{11} = gC^2\left(1+2C^2\right)\Gamma(\omega) \tag{21d}$$

The element $\rho_{22}$ may be found from the probability conservation low $\rho_{22} = 1 - \rho_{11}$ [1].

### 3. Dynamics of quantum antenna: qualitative analysis

The strong coupling regime corresponds to the condition $\Omega_R \gg \Delta$, which is equal to $g \approx 1$. We will analyze for simplicity antenna dynamics in the regime of exact resonance (zero detuning). In this case the approximation $\Gamma(\omega \pm 2\nu) \approx \Gamma(\omega)$ becomes appropriate, and $\gamma_{12} \approx 3\Gamma(\omega)/2$, $\gamma_{21} \approx \Gamma(\omega)/2$ $\gamma_{11} \approx \gamma_{22} \approx \Gamma(\omega)$. Therefore, the system (9)-(11) will be simplified to

$$\dot{\rho}_{11} = -\frac{\Gamma(\omega)}{2}\rho_{11} + \frac{\Gamma(\omega)}{2} \tag{22a}$$

$$\dot{\rho}_{12} = -i\left(2\nu - i\frac{3}{4}\Gamma(\omega)\right)\rho_{12} - \frac{\Gamma(\omega)}{4}\rho_{21} + \frac{\Gamma(\omega)}{2} \tag{22b}$$

$$\dot{\rho}_{21} = i\left(2\nu + i\frac{3}{4}\Gamma(\omega)\right)\rho_{21} - \frac{\Gamma(\omega)}{4}\rho_{12} + \frac{\Gamma(\omega)}{2} \tag{22c}$$

The system (22a)-(22c) is equal to the equations of resonant fluorescence [1]. The temporal evolution described by this system goes to the steady state defined by the condition $\dot{\rho}_{\alpha\beta} = 0$. For the strong fields such that $\Omega_R \gg \Gamma$ the last two terms in [1] have been neglected. As a result, the steady state values $\rho_{11}(\infty) = \rho_{22}(\infty) = 1/2$ and $\rho_{12}(\infty) = \rho_{21}(\infty) = 0$. We are keeping these terms and obtain the steady state as

$$\rho(\infty) = \begin{pmatrix} 1/2 & \rho_{12}(\infty) \\ \rho_{12}^*(\infty) & 1/2 \end{pmatrix} \tag{23}$$

where

$$\rho_{12}(\infty) = \frac{\Gamma(\omega)\left(2i\nu - \dfrac{\Gamma(\omega)}{2}\right)}{2\left(4\nu^2 - \dfrac{\Gamma^2(\omega)}{2}\right)} \tag{24}$$

As it leads from (20), the antenna emission has not vanished, while the emission properties are independent on its initial state.

## 4. Radiation properties of quantum antenna

In this section we will consider the power radiation pattern of quantum antenna in the strong coupling regime. The far field emitted by antenna [1] presented as a superposition of the partial supports produced by elementary dipoles induced at the antenna surface. The positive-frequency part of field operator produced by the single dipole quantum emitter [1] in spherical system is given by $\hat{\mathbf{E}}^{(+)}(\mathbf{r},t) = \mathbf{e}_x \omega_0^2 d_{ab} \hat{\sigma}^-(t-r/c)/4\pi\varepsilon_0 c^2 r$, where $r$ is the spherical radial coordinate (dipole assumed to be placed at the origin). The correspondent operator of the antenna field is

$$\hat{\mathbf{E}}^{(+)}(\mathbf{r},t) = \mathbf{e}_x \frac{\omega_0^2 d_{ab}}{4\pi\varepsilon_0 c^2} \int_{-l/2}^{l/2} \frac{1}{|\mathbf{r}-\mathbf{r}'(x')|} f(x') \hat{\sigma}^-\left(t - \frac{|\mathbf{r}-\mathbf{r}'(x')|}{c}\right) dx' \tag{25}$$

The normally ordered operator of intensity in the far field zone is given by

$$\hat{\mathbf{E}}^{(-)}(\mathbf{r},t) \cdot \hat{\mathbf{E}}^{(+)}(\mathbf{r},t) = \mathbf{e}_x \frac{\omega_0^4 d_{ab}^2}{(4\pi\varepsilon_0)^2 c^4} \int_{-l/2}^{l/2} \int_{-l/2}^{l/2} \frac{f(x)}{|\mathbf{r}-\mathbf{r}'(x)|} \frac{f^*(x')}{|\mathbf{r}-\mathbf{r}'(x')|} \hat{\sigma}^+\left(t - \frac{|\mathbf{r}-\mathbf{r}'(x)|}{c}\right) \cdot \hat{\sigma}^-\left(t - \frac{|\mathbf{r}-\mathbf{r}'(x')|}{c}\right) dx' dx \tag{26}$$

The observable value of intensity expressed through the two-time correlation function of polarization [1]. It reads

$$\left\langle \hat{\mathbf{E}}^{(-)}(\mathbf{r},t) \cdot \hat{\mathbf{E}}^{(+)}(\mathbf{r},t) \right\rangle = \mathbf{e}_x \frac{\omega_0^4 d_{ab}^2}{(4\pi\varepsilon_0)^2 c^4} \int_{-l/2}^{l/2} \int_{-l/2}^{l/2} \frac{f(x)}{|\mathbf{r}-\mathbf{r}'(x)|} \frac{f^*(x')}{|\mathbf{r}-\mathbf{r}'(x')|} K(x,x';t) dx' dx \tag{27}$$

where

$$K(x,x;t) = \left\langle \hat{\sigma}^+\left(t - \frac{|\mathbf{r}-\mathbf{r}'(x)|}{c}\right) \cdot \hat{\sigma}^-\left(t - \frac{|\mathbf{r}-\mathbf{r}'(x')|}{c}\right) \right\rangle \tag{28}$$

We will consider the steady state of antenna given by equations (23),(24). It is stationary process, for which the correlation function is time-independent. Thus, equation (28) may be rewritten as

$$K(x,x') = \left\langle \hat{\sigma}^+(0) \cdot \hat{\sigma}^-\left(\frac{|\mathbf{r}-\mathbf{r}'(x)|-|\mathbf{r}-\mathbf{r}'(x')|}{c}\right) \right\rangle \tag{29}$$

The correlation function (29) is equal to the correlation function of resonance fluorescence in detail considered in [1]. An approximate relation for steady state reads

$$\left\langle \hat{\sigma}^+(0) \cdot \hat{\sigma}^-(\tau) \right\rangle_{ss} = \frac{1}{4}\left( e^{-\frac{\Gamma}{2}\tau} + \frac{1}{2} e^{-\frac{3\Gamma}{4}\tau} \cdot e^{-i\Omega_R \tau} + \frac{1}{2} e^{-\frac{3\Gamma}{4}\tau} \cdot e^{i\Omega_R \tau} \right) e^{-i\omega\tau} \tag{30}$$

The time shift in antenna is stipulated by the phase shift of radiation from different points and given by $\tau = (|\mathbf{r}-\mathbf{r}'(x)| - |\mathbf{r}-\mathbf{r}'(x')|)/c$. For simplifying the integration in (27), we will use the conventional assumptions to macroscopic antenna theory [14]:

$$|\mathbf{r}-\mathbf{r}'(x')| \approx R \tag{31a}$$

$$|\mathbf{r}-\mathbf{r}'(x')| \approx R - x'\cos\theta, \tag{31b}$$

for amplitude and phase factors, respectively, where $R, \theta$ are coordinates for the spherical system with origin at the antenna center (exponentially attenuated factors in (30) are related to the amplitude ones and approximated accordingly (31a)). As a result, the intensity of radiation may be presented in terms of radiation pattern, conventional for antenna theory [14]. It reads

$$\left\langle \hat{\mathbf{E}}^{(-)}(\mathbf{r},t) \cdot \hat{\mathbf{E}}^{(+)}(\mathbf{r},t) \right\rangle_{SS} \approx \mathbf{e}_x \frac{\omega_0^4 d_{ab}^2}{(4\pi\varepsilon_0)^2 c^4} \frac{1}{R^2} \cos^2\theta \cdot \xi(\theta) \tag{32}$$

where

$$\xi(\theta) = 2\,\mathrm{Re} \int_{-l/2}^{l/2} dx \cdot f(x) \int_{x}^{l/2} f^*(x') \left\langle \hat{\sigma}^+(0)\hat{\sigma}^-\left(\frac{x'-x}{c}\cos\theta\right) \right\rangle_{SS} dx' \tag{33}$$

($\theta < |\pi/2|$).

For illustrating the qualitative properties of radiation pattern we consider the simplest model of perfect linear antenna [14], which quantum envelope has the constant spatial amplitude and the linear phase distribution $f(x) = e^{ik\varphi x}$, where $\varphi$ is a phase shift per unit length. The integration in (33) gives

$$\xi(\theta) = \frac{1}{2}\left\{ \left(\frac{\sin\Psi}{\Psi}\right)^2 + \frac{1}{2}\left(\frac{\sin\Psi_+}{\Psi_+}\right)^2 + \frac{1}{2}\left(\frac{\sin\Psi_-}{\Psi_-}\right)^2 \right\} \tag{34}$$

with $\Psi = kl(\cos\theta - \varphi)/2$, $\Psi_\pm = \Psi \pm (\Omega_R/\omega)\cos\theta$.

Let us consider now the angular pattern of power spectrum of quantum antenna for the strong-field limit ($\Omega_R \gg \Gamma/4$). Following Wiener-Khintchine theorem, it is given in terms of the two-time correlation function by the field in the far zone by

$$S(\omega; R, \theta) = \frac{1}{\pi}\mathrm{Re}\int_0^\infty \left\langle \hat{\mathbf{E}}^{(-)}(\mathbf{r},t) \cdot \hat{\mathbf{E}}^{(+)}(\mathbf{r},t+\tau) \right\rangle_{SS} e^{i\omega\tau} d\tau =$$
$$\approx \frac{1}{\pi}\frac{\omega_0^4 d_{ab}^2}{(4\pi\varepsilon_0)^2 c^4} \frac{1}{R^2} \cos^2\theta \cdot \mathrm{Re}\int_0^\infty \xi(\tau;\theta) e^{i\omega\tau} d\tau \tag{35}$$

where

$$\xi(\tau;\theta) = \int_{-l/2}^{l/2} dx \cdot f(x) \int_{-l/2}^{l/2} f^*(x') \left\langle \hat{\sigma}^+(0)\hat{\sigma}^-\left(\tau + \frac{x'-x}{c}\cos\theta\right) \right\rangle_{SS} dx' \tag{36}$$

Making the integrations in (35), (36), we obtain the final result in the form

$$S(\omega;R,\theta) = \frac{k^4 d_{ab}^2 \sin^2\theta}{4\pi(4\pi\varepsilon_0)^2 R^2}\Gamma \times$$

$$\left\{\frac{[\sin(\Psi)/\Psi]^2}{(\omega-\omega_0)^2+(\Gamma/2)^2} + \frac{(3/4)[\sin(\Psi_+)/\Psi_+]^2}{(\omega-\omega_0-\Omega_R)^2+(3\Gamma/4)^2} + \frac{(3/4)[\sin(\Psi_-)/\Psi_-]^2}{(\omega-\omega_0+\Omega_R)^2+(3\Gamma/4)^2}\right\} \quad (37)$$

Equation (37) allows analyzing the qualitative behavior of radiation pattern, some of which aspects is seen to be in agreement with theory of resonance fluorescence [1]. The total radiation pattern represents the sum of three elementary patterns $(\sin\Psi/\Psi)^2$ of the form of perfect wire antenna [14], centered at the three different angles $\theta = \arccos(2\varphi/kl)$, $\theta_\pm = \arccos[2\varphi(1\pm\Omega_R/\omega)/kl]$. The weight of the elementary pattern is defined by the correspondent resonant line centered at one of the frequencies $\omega=\omega_0$, $\omega=\omega_0+\Omega_R$, $\omega=\omega_0-\Omega_R$. Such form of frequency spectrum corresponds to Mollow triplet in resonance fluorescence for the case of strong driving field [1]. The element of novelty in the antenna is that radiation of each triplet-line occurs in its own direction. The value of angle splitting depends on the driven field, phase shift of the quantum envelope along antenna and its length compared with the wavelength.

The trends of mentioned behavior are illustrated by Figures 2-4. The elementary radiation patterns of the lines $\omega=\omega_0$ (colored red), $\omega=\omega_0+\Omega_R$ (colored blue) and $\omega=\omega_0-\Omega_R$ (colored green) in logarithmic scale for different values of driven parameters have been shown at this Figures. The patterns are in the ratio 2:1:1 accordingly with the ratio of integrated intensities in the peaks. It means that the every elementary pattern corresponds to the signal formed by the broadened spectral line with the width $\Gamma/2, 3\Gamma/4, 3\Gamma/4$, respectively. Figure 1a approximately corresponds to the typical pattern of the perfect wire. The line $\omega=\omega_0$ have been not seen, because of for such small values of driven field it runs into the line $\omega=\omega_0+\Omega_R$. It is observed rather small difference between the powers of lines, but no their angle splitting. Increasing of Rabi-frequency (Figure 2b) keeps number of lobes to be the same, but leads to the angle splitting of rather small value for the main lobe.

The patterns become more and more jagged with the length increasing (Figures 3,4). Simultaneously, effect of pattern separation increases and even the absolute separation becomes reachable (the angle of maximum for the main lobe of one pattern coincides with the minimum for another one). The typical radiation patterns of wire antennas are highly dependent on the value of phase shift. The value $\varphi=\varphi_{cr}$ defines the regime of axial radiation. For the phase shift exceeding the critical value $\varphi>\varphi_{cr}$, the main lobe removes to the invisible region [14] and becomes not related to the observable angles $\theta$. Therefore, the antenna emission is completely formed as a superposition of side lobes, which are incoherent. As a result, the emitted power decreases, which makes this regime to be not suitable for a lot of applications. For the classical wire antennas $\varphi_{cr}=1$ [14]. In our case we have the own critical shift for every partial diagram: $\varphi_{cr}=1$ for central line and $\varphi_{cr}=1/\left(1\pm\dfrac{\Omega_R}{\omega}\right)$ for two side lines. The last critical values are dependent on Rabi-frequency and become controllable via the adiabatic field variance. For example, the right-hand line may be removed to the invisible region while two others left in the visible one. This mechanism manifests itself in the strong decreasing of the emitted power at the line $\omega=\omega_0+\Omega_R$ compared with two others for rather large Rabi-frequency (Figures 3a, 4a).

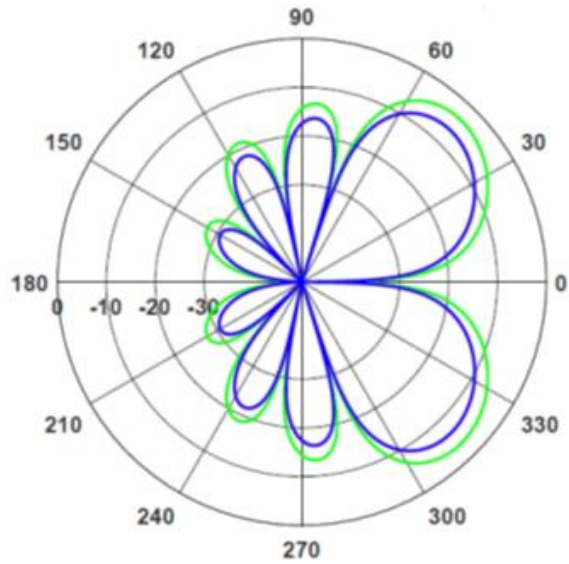

(a)

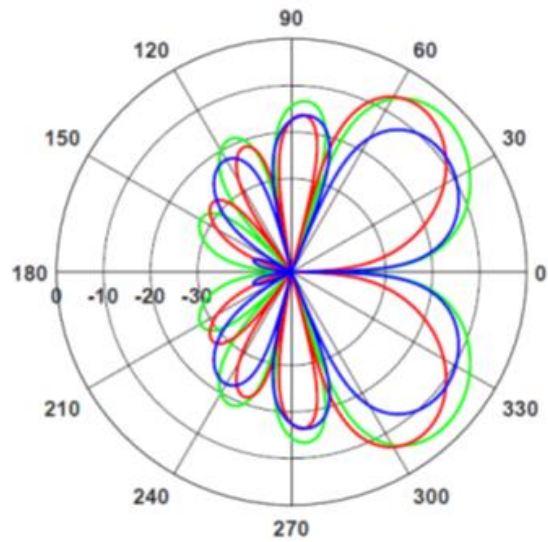

(b)

**Figure 2**. The elementary radiation patterns of the lines $\omega = \omega_0$ (colored red), $\omega = \omega_0 + \Omega_R$ (colored blue) and $\omega = \omega_0 - \Omega_R$ (colored green) for different values of Rabi-frequency. The patterns are in the ratio 2:1:1 accordantly with the ratio of integrated intensities in the peaks. The patterns presented in logarithmic scale. $kl/2 = 2\pi$, $\varphi = 0.8$. (a) $\Omega_R = 0.001\omega$. (b) $\Omega_R = 0.2\omega$

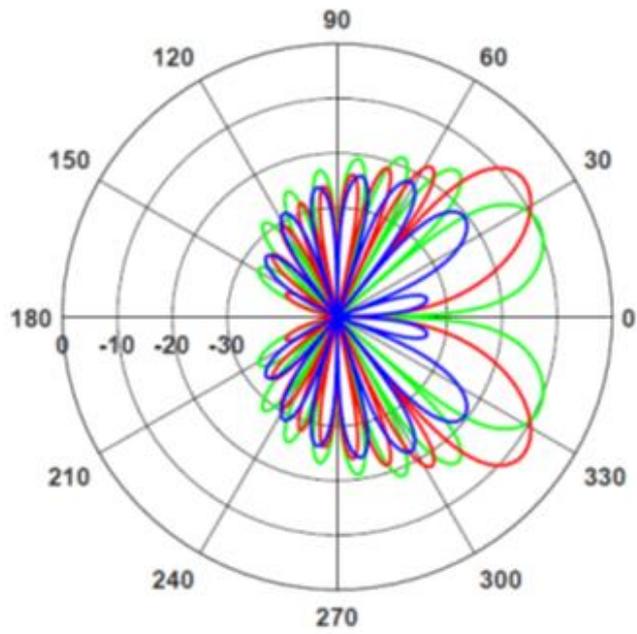

(a)

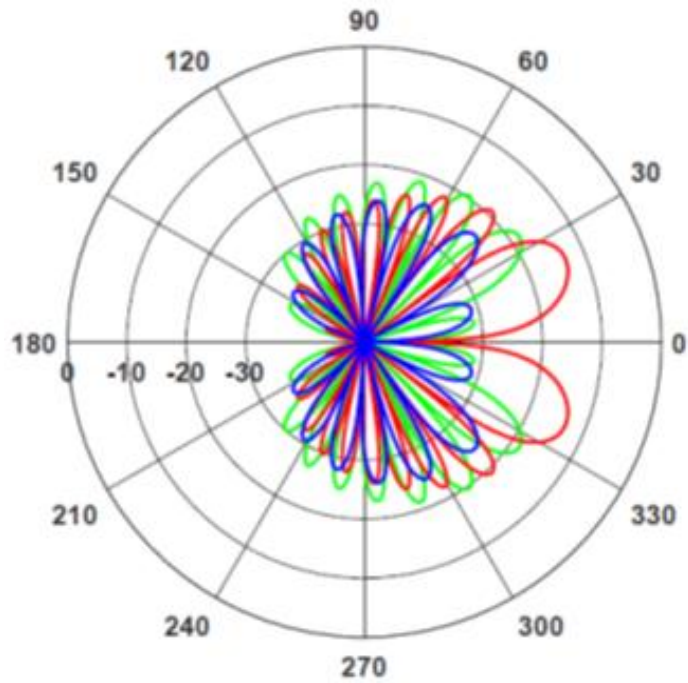

(b)

**Figure 3** The elementary radiation patterns of the lines $\omega = \omega_0$ (colored red), $\omega = \omega_0 + \Omega_R$ (colored blue) and $\omega = \omega_0 - \Omega_R$ (colored green) in logarithmic scale for different values of phase shift. $kl/2 = 4\pi$, $\Omega_R = 0.2\omega$. (**a**) $\varphi = 1.0$. (**b**) $\varphi = 1.2$.

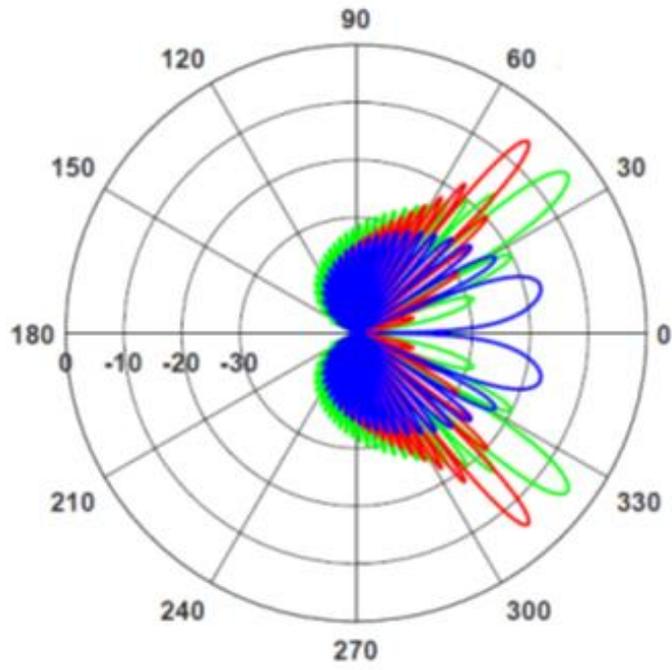

(a)

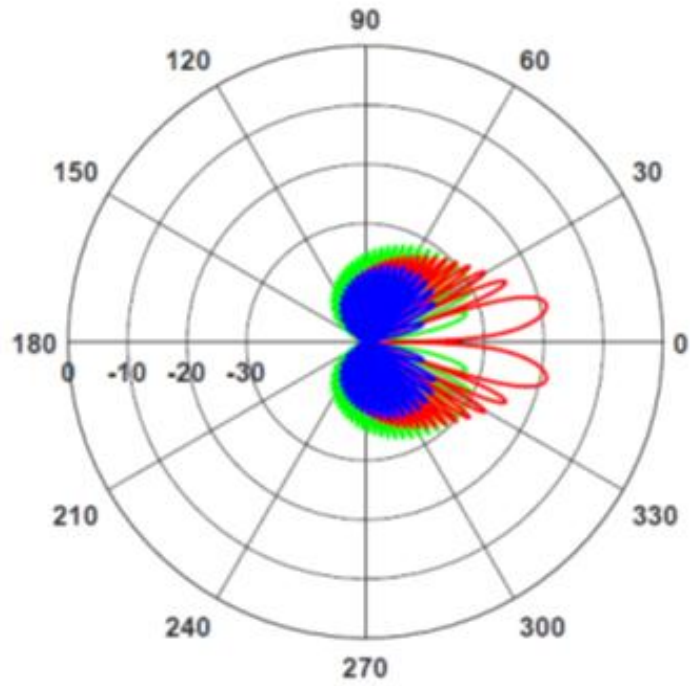

(b)

**Figure 4.** The elementary radiation patterns of the lines $\omega = \omega_0$ (colored red), $\omega = \omega_0 + \Omega_R$ (colored blue) and $\omega = \omega_0 - \Omega_R$ (colored green) in logarithmic scale for different values of phase shift. $kl/2 = 15\pi$, $\Omega_R = 0.2\omega$. (**a**) $\varphi = 0.8$. (**b**) $\varphi = 1.2$.

## 5. Conclusion

In summary, we developed the model of quantum antenna in the strong coupling regime basing on the general theory of open quantum systems [1,33]. The far-field zone of antenna radiation is considered as the thermal photonic reservoir. We formulated and solved the master equation with Lindblad-terms related to the energy losses via antenna emission. The general concept was applied to the wire antenna with Fermionic type of excitation. Spectral density of power in far-field zone was calculated. It is shown, that the strong coupling regime dramatically changes the radiation pattern compared with the macroscopic antennas and optical nanoantennas of different well-known types. The calculated radiation pattern consists of three components, each corresponding to the resonance line in the Mollow triplet and turned one with respect to another. The value of turn strongly depends on the geometric parameters of antenna, energy spectra of used materials and the value of coupling (Rabi-frequency). It opens a new ways for high-effective electric control of antennas characteristics for using in different nano-photonic applications. On the other hand, the influence of EM-field on the pattern should be accounted from the point of view of electromagnetic compatibility in nanoscale [37,38].